\documentclass[superscriptaddress,
reprint,
 amsmath,amssymb,
 aps,
]{revtex4-1}

\usepackage{graphicx}
\usepackage{dcolumn}
\usepackage{bm}

\begin{document}

\preprint{APS/123-QED}

\title{Real photons produced from photoproduction in $pp$ collisions}

\author{FU Yong-Ping }
\affiliation{Department of Physics, Yunnan University, Kunming, 650091, China}
\affiliation{Department of Physics and Mathematics, Lincang Teachers College, Lincang, 677000, China}
\author{LI Yun-De }
\email{Corresponding author \emph{E-mail address:} yndxlyd@163.com}
\affiliation{Department of Physics, Yunnan University, Kunming, 650091, China}


\begin{abstract}
We calculate the production of real photons originating from the
photoproduction in relativistic $pp$ collisions.
The Weizs$\ddot{\mathrm{a}}$cker-Williams approximation
in the photoproduction is considered. Numerical results agree with the experimental data from
Relativistic Heavy Ion Collider (RHIC) and Large
Hadron Collider (LHC). We find that the modification of the photoproduction is more prominent in large
transverse momentum region.
\\PACS: 25.20.Lj, 12.39.St, 12.38.Bx
\end{abstract}

\maketitle

Since photons do not participate in the strong interaction, the observation of prompt photons from the hadron collisions is a
good method to test the perturbative Quantum Chromodynamics (pQCD). The application of pQCD for various hard
scattering processes of the prompt photon production has been investigated
\cite{1,12.1,19.1,19.2,19.3,19.4}. It is necessary to identify different prompt photon sources in relativistic hadronic collisions. The hard
scattering of partons is a well-known source of large transverse
momentum ($P_{T}$) photons in high energy hadronic collisions. The
prompt photons include direct photons and fragmentation photons \cite{2,3,15}. The direct photons are those produced by the Compton scattering
($qg\rightarrow q\gamma$) and the annihilation of two partons
($q\bar{q}\rightarrow g\gamma$). The fragmentation photons are those produced by the bremsstrahlung emitted from
final state partons $\left(ab\rightarrow (c\rightarrow
x\gamma)d\right)$. Previous works have studied the next-to-leading-order (NLO) contributions of the prompt photon production \cite{12,12.01},
but the contributions of the photoproduction (higher-order) in the $pp$ collisions are not researched clearly.

The photoproduction
is an important concept in the $ep$ deep inelastic scattering (DIS)
at Hadron Electron Ring Accelerator (HERA) \cite{21,22,23}. In
the $ep$ DIS, the electron interacts with the partons of the proton by exchanging a photon,
a high energy photon emitted from the incident
electron ($e\rightarrow e\gamma$) directly interacts with the proton by the interaction of
$\gamma p\rightarrow $X (photoproduction). Besides, the Heisenberg's uncertainty principle
allows the high energy photon for a short time to fluctuate into a quark-antiquark pair which then interacts
with the partons of the proton (resolved photoproduction). In such interactions the photons emitted from the incident
electron can be regarded as an extended object
consisting of quarks and also gluons. The photons
are the so-called resolved photons ($\gamma_{resolved}$).

In this work, we study the large $P_{T}$ photons from the photoproduction mechanism by considering the Weizs$\ddot{\mathrm{a}}$cker-Williams approximation in relativistic $pp$ collisions. The charged partons of the incident proton can emit
high energy photons ($q\rightarrow q\gamma$) and resolved photons ($q\rightarrow q\gamma_{resolved}$) in relativistic $pp$ collisions \cite{24.1,24.2,24,25.01,25}, then the high energy photons
or resolved photons interact with the partons of another proton by the
subprocesses of $q\gamma\rightarrow q\gamma$,
$q_{\gamma}\bar{q}\rightarrow g\gamma$, $\bar{q}_{\gamma}q\rightarrow g\gamma$, $q_{\gamma}g\rightarrow
q\gamma$ and $g_{\gamma}q\rightarrow q\gamma$, here
$q_{\gamma}$($g_{\gamma}$) denotes the parton of the resolved
photon.

The prompt photons
produced by the leading-order (LO) QCD Compton scattering, annihilation, and
bremsstrahlung from final state partons have been discussed
by previous works \cite{1,15,12}. In this Letter we focus on the
higher-order modification: the photoproduction. In the
photoproduction, a parton $q_{a}$ of an incident proton $A$ emits a
high energy photon ($q_{a}\rightarrow q_{a}\gamma$), then the photon interacts with a parton $q_{b}$ of
another incident proton $B$ by the interaction of
$\gamma p\rightarrow \gamma X$. The momentum transfer $Q^{2}$ in the deep inelastic collisions is large, so the perturbative factorization is applicable. The large $P_{T}$ photons produced
by photoproduction (pho.) satisfy the
following invariant cross section \cite{24.1,24.2}
\begin{eqnarray}
E\frac{d\sigma_{pho.}}{d^{3}P}\label{eq008}&=&\frac{2}{\pi}\int^{1}_{x_{amin}}
dx_{a}\int^{1}_{x_{bmin}}
dx_{b}F_{q_{a}/A}(x_{a},Q^{2})\nonumber\\[1mm]
&&\times F_{q_{b}/B}(x_{b},Q^{2})f_{\gamma/q_{a}}(z_{a})
\frac{x_{a}x_{b}z_{a}}{x_{a}x_{b}-x_{a}x_{2}}
\nonumber\\[1mm]
&&\times \frac{d\hat{\sigma}}{d\hat{t}}(x_{a},x_{b},z_{a},P_{T}),
\end{eqnarray}
where $x_{a}$ and $x_{b}$ are the momentum fractions of partons, $F_{q_{a}/A}(x_{a},Q^{2})$ and $F_{q_{b}/B}(x_{b},Q^{2})$ are the
parton distribution of the proton \cite{5}, we choose $Q^{2}=4P_{T}^{2}$ in the
distribution \cite{2}. $f_{\gamma/q_{a}}(z_{a})$ is the
photon spectrum from the quark $q_{a}$. The cross section of LO
subprocesses $d\hat{\sigma}/d\hat{t}$ is QED Compton
scattering $q_{b}\gamma\rightarrow q\gamma$. The minimum values of $x_{a}$ and $x_{b}$ in the
integral are
\begin{eqnarray}
x_{amin}=\frac{x_{1}}{1-x_{2}},
\end{eqnarray}
and
\begin{eqnarray}
x_{bmin}=\frac{x_{a}x_{2}}{x_{a}-x_{1}}.
\end{eqnarray}
The momentum fraction $z_{a}$ of the photon emitted from the quark
is
\begin{eqnarray}
z_{a}=\frac{x_{b}x_{1}}{x_{a}x_{b}-x_{a}x_{2}},
\end{eqnarray}
where the variables are
\begin{eqnarray}
x_{1}=\frac{1}{2}x_{T}e^{y},
\end{eqnarray}
and
\begin{eqnarray}
x_{2}=\frac{1}{2}x_{T}e^{-y},
\end{eqnarray}
and
\begin{eqnarray}
x_{T}=\frac{2P_{T}}{\sqrt{s_{NN}}},
\end{eqnarray}
here $\sqrt{s_{NN}}$ is the total energy in the
center-of-mass system, and $y$ is the rapidity of the real photons.

The pQCD requests the QCD momentum transfer $Q^{2}>$ 1 GeV$^{2}$ \cite{25}. In the photoproduction high energy photons are bremsstrahlung from the charged particles,
this is a pure QED process. Since the QCD coupling parameter $\alpha_{s}(Q^{2})$ does not depend on the
QED momentum transfer $q^{2}$, the Weizs$\ddot{\mathrm{a}}$cker-Williams approximation of the photoproduction is still valid for the charged partons of protons \cite{25.01,25}. Therefore the photon spectrum from the quark can be described by the Weizs$\ddot{\mathrm{a}}$cker-Williams distribution
\begin{eqnarray}
&&f_{\gamma/q_{a}}(z_{a})\label{pho}=\int^{q^{2}_{max}}_{q^{2}_{min}}dq^{2}\frac{dN_{\gamma}}{dzdq^{2}}=\nonumber\\[1mm]
&&\times
\frac{e_{q}^{2}\alpha P_{q\rightarrow \gamma}(z_{a})}{2\pi}
\left[\ln\left(\frac{(\sqrt{s_{NN}})^{2}}{4m^{2}_{q}}\right)+\ln\left(\frac{x_{a}x_{b}(1-z_{a})}{z_{a}}\right)\right]\nonumber\\[1mm]
&&-\frac{e_{q}^{2}\alpha }{\pi}\left(\frac{1-z_{a}}{z_{a}}-\frac{4m_{q}^{2}}{x_{a}x_{b}s_{NN}}\right),
\end{eqnarray}
where $e_{q}$ is the charge of the quark, $\alpha$ is the
electromagnetic coupling parameter, $P_{q\rightarrow \gamma}(z_{a})=\left(1+(1-z_{a})^{2}\right)/z_{a}$ is the split function. The maximum momentum transfer $q_{max}^{2}$ is determined by the
experimental acceptance \cite{21}. According to \cite{25} we choose
$q_{max}^{2}$ as $\hat{s}/4$. By considering the
Weizs$\ddot{\mathrm{a}}$cker-Williams approximation, the kinematic
limit of minimum momentum transfer $q_{min}^{2}=(m_{q}z_{a})^{2}/(1-z_{a})$ is made such that the
photon is close to being on its mass shell \cite{22,23}. Here $m_{q}$ is the mass of the quark. The photoproduction contain higher-order QED
coupling parameters, but the collision energies at RHIC ($\sqrt{s_{NN}}$
= 200 GeV) and LHC ($\sqrt{s_{NN}}$ = 5500 GeV) are large enough.
Therefore the values of the term $\ln\left((\sqrt{s_{NN}})^{2}/4m^{2}_{q}\right)$ in Eq. (\ref{pho}) enhance the
modification of the photoproduction \cite{24.1}.

The QED Compton process ($q_{b}\gamma\rightarrow q\gamma$) can be written
as \cite{2}
\begin{eqnarray}
\frac{d\hat{\sigma}}{d\hat{t}}(q\gamma\rightarrow
q\gamma)\label{QED1}=\frac{\pi\alpha^{2}e_{q}^{4}}{\hat{s}^{2}}2\left(-\frac{\hat{t}}{\hat{s}}-\frac{\hat{s}}{\hat{t}}\right),
\end{eqnarray}
where the Mandelstam variables are
\begin{eqnarray}
\hat{s}=x_{a}x_{b}z_{a}s_{NN},
\end{eqnarray}
and
\begin{eqnarray}
\hat{t}=-x_{a}x_{2}z_{a}s_{NN}.
\end{eqnarray}

\begin{figure}[b]
\includegraphics[scale=0.75]{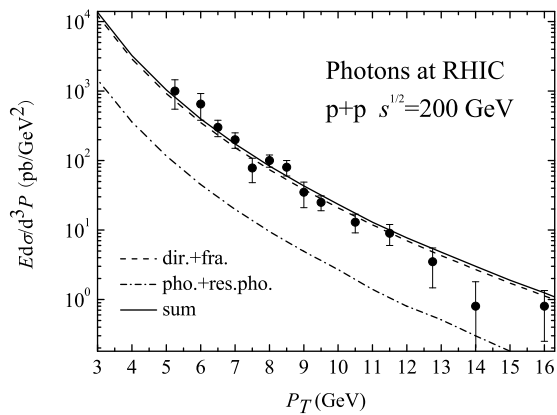}
\caption{The cross section of the real photon production in relativistic $pp$
collisions at RHIC. The dash line is the sum of direct photons and
fragmentation photons \cite{24.1}. The dash dot line represents photons produced
by the photoproduction and resolved photoproduction. The solid line is
the sum of direct photons, fragmentation photons and photons
from the photoproduction and resolved photoproduction. Data points are from PHENIX at RHIC \cite{12.1}.\label{fig1}}
\end{figure}

The high energy photon emitted from the quark of the proton can fluctuate into a
parton anti-parton pair in a short period of time $\triangle t$. Besides the (anti-)quarks may emit gluons in the time $\triangle t$. If,
during such a fluctuation, one of the partons from the fluctuated photon interacts with
partons of another incident proton. In such interaction the high energy
photon is resolved \cite{21,22,23}. In the resolved photoproduction, the parton $q_{a}$ of the incident proton $A$ emits a high
energy resolved photon ($q_{a}\rightarrow q_{a}\gamma_{resolved}$), then the parton of the resolved photon ($\gamma_{resolved}\rightarrow$$q_{a'}\bar{q}_{a'}$ or $q_{a'}\bar{q}_{a'}g_{a'}$)
interacts with the parton ($q_{b}$ or $g_{b}$) of another incident proton $B$ by the
interaction of $\gamma_{resolved} p\rightarrow \gamma X$. The corresponding invariant cross section of large $P_{T}$
 photons
produced by resolved photoproduction (res. pho.) can be
written as \cite{24.1,24.2}
\begin{eqnarray}
E\frac{d\sigma_{res.pho.}}{d^{3}P}\label{eq009}&=&\frac{2}{\pi}\int^{1}_{x_{amin}}
dx_{a}\int^{1}_{x_{bmin}} dx_{b}\int^{1}_{z_{a' min}} dz_{a'}
\nonumber\\[1mm]
&&\times F_{q_{a}/A}(x_{a},Q^{2})F_{q_{b}/B}(x_{b},Q^{2})f_{\gamma/q_{a}}(z_{a})\nonumber\\[1mm]
&&\times
F_{q_{a'}/\gamma}(z_{a'},Q^{2})\frac{x_{a}x_{b}z_{a}z_{a'}}{x_{a}x_{b}z_{a'}-x_{a}z_{a'}x_{2}}
\nonumber\\[1mm]
&&\times
\frac{d\hat{\sigma}}{d\hat{t}}(x_{a},x_{b},z_{a},z_{a'},P_{T}),
\end{eqnarray}
where $z_{a'}$ is the momentum fraction of partons from the resolved
photon, and $F_{q_{a'}/\gamma}(z_{a'},Q^{2})$ is the parton
distribution of the resolved photon \cite{7}. The minimum values of
momentum fractions are
\begin{eqnarray}
x_{amin}=\frac{x_{1}}{1-x_{2}},
\end{eqnarray}
and
\begin{eqnarray}
x_{bmin}=\frac{x_{a}x_{2}}{x_{a}-x_{1}},
\end{eqnarray}
and
\begin{eqnarray}
z_{a'min}=\frac{x_{b}x_{1}}{x_{a}x_{b}-x_{a}x_{2}}.
\end{eqnarray}

The variable $z_{a}$ of the resolved photon is given by
\begin{eqnarray}
z_{a}=\frac{x_{b}x_{1}}{x_{a}x_{b}z_{a'}-x_{a}z_{a'}x_{2}}.
\end{eqnarray}
In the resolved photoproduction, the photon spectrum from
the quark $q_{a}$ can be written as
\begin{eqnarray}
&&f_{\gamma/q_{a}}(z_{a})\label{rs}=\nonumber\\[1mm]
&&\times
\frac{e_{q}^{2}\alpha P_{q\rightarrow \gamma}(z_{a})}{2\pi}\!\!
\left[\ln\left(\frac{(\sqrt{s_{NN}})^{2}}{4m^{2}_{q}}\right)\!\!+\!\!\ln\left(\frac{x_{a}x_{b}z_{a'}(1-z_{a})}{z_{a}}\right)\right]\nonumber\\[1mm]
&&-\frac{e_{q}^{2}\alpha }{\pi}\left(\frac{1-z_{a}}{z_{a}}-\frac{4m_{q}^{2}}{x_{a}x_{b}z_{a'}s_{NN}}\right).
\end{eqnarray}

\begin{figure}[t]
\includegraphics[scale=0.75]{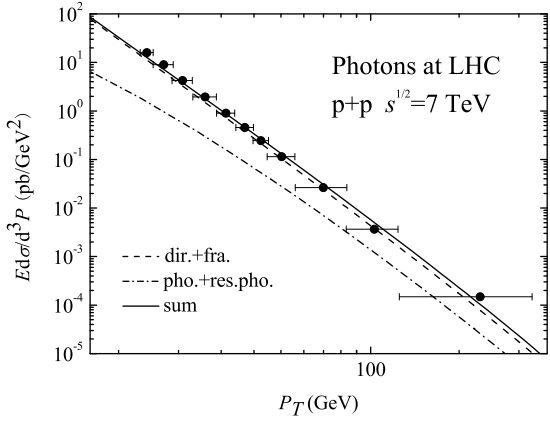}
\caption{Same as Fig.\ref{fig1} but for $pp$ $\sqrt{s}$=7 TeV
collisions at LHC. Data points are from CMS at LHC \cite{9.01}.\label{fig2}}
\end{figure}

The cross section $d\hat{\sigma}/d\hat{t}$ of the annihilation and
Compton scattering ($q_{a'}\bar{q}_{b}\rightarrow g \gamma$, $\bar{q}_{a'}q_{b}\rightarrow g \gamma$,
$q_{a'}g_{b}\rightarrow q
 \gamma$, $g_{a'}q_{b}\rightarrow
q
 \gamma$) are given by \cite{1,2}
\begin{eqnarray}
\frac{d\hat{\sigma}}{d\hat{t}}(q\bar{q}\rightarrow
g\gamma)\label{sub1}=\frac{\pi\alpha\alpha_{s}e_{q}^{2}}{\hat{s}^{2}}\frac{8}{9}\left(\frac{\hat{u}}{\hat{t}}+\frac{\hat{t}}{\hat{u}}\right),
 \end{eqnarray}
and
\begin{eqnarray}
\frac{d\hat{\sigma}}{d\hat{t}}(qg\rightarrow
q\gamma)\label{sub2}=\frac{\pi\alpha\alpha_{s}e_{q}^{2}}{\hat{s}^{2}}\frac{1}{3}\left(-\frac{\hat{t}}{\hat{s}}-\frac{\hat{s}}{\hat{t}}\right),
 \end{eqnarray}
where the strong coupling constant is
\begin{eqnarray}
\alpha_{s}=\frac{12\pi}{(33-2n_{f})\ln(Q^{2}/\Lambda^{2})},
\end{eqnarray}
the large momentum variable is $Q^{2}=4P_{T}^{2}$ and the momentum scale is $\Lambda_{LO}$= 0.2 GeV \cite{2}. $n_{f}$ is the flavor number. The
Mandelstam variables in resolved photoproduction are
\begin{eqnarray}
\hat{s}=x_{a}x_{b}z_{a}z_{a'}s_{NN},
\end{eqnarray}
and
\begin{eqnarray}
\hat{t}=-x_{a}x_{2}z_{a}z_{a'}s_{NN},
\end{eqnarray}
and
\begin{eqnarray}
\hat{u}=-x_{b}x_{1}s_{NN}.
\end{eqnarray}

In Fig.\ref{fig1} and \ref{fig2} we plot the contribution of prompt photons and photons
produced from photoproductions for relativistic $pp$
collisions at RHIC and LHC energies. Except the leading QCD Compton and annihilation subprocesses, the pure QED annihilation $q\bar{q}\rightarrow\gamma\gamma$ (res.pho.) and QCD-induced gluon-photon coupling $g\gamma\rightarrow g\gamma$ (pho.), $g\gamma\rightarrow \gamma\gamma$ (pho.), $gg\rightarrow g\gamma$ (res.pho.) and $gg\rightarrow \gamma\gamma$ (res.pho.) \cite{23.01}
are considered in the subprocesses. The correction factors to take into account the NLO effect
of subprocesses of direct photons and fragmentation photons are $K_{dir. \gamma}$ and $K_{fra. \gamma}$. We evaluate these $P_{T}$-dependent $K$ factors using the
numerical program from Aurenche et al \cite{12.01}. The authors of Ref. \cite{12} have obtained $K_{dir. \gamma}$(10 GeV)$\sim$ 1.5 for RHIC and LHC
and $K_{fra. \gamma}$(10 GeV)$\sim$ 1.8 at RHIC and 1.4 at LHC.

In Fig.\ref{fig1} the spectra
of the photoproduction and resolved photoproduction (dash dot line) are compared with the spectra of direct photons
and fragmentation photons (dash line). The numerical results show that the modification
of photoproductions is not prominent in the relatively small $P_{T}$ region. The modification of the photoproductions is
$\left(Ed\sigma_{ pho.\!+\!res.pho.} /d^{3}P\right)/\left(Ed\sigma_{dir.\!+\!fra.} /d^{3}P\right)\sim$ 12$\%$ in the region of 5 GeV $<P_{T}<$ 16 GeV  at RHIC energies.
However, the photoproductions start playing an interesting
role in large $P_{T}$ region at LHC energies. The modification
in the region of 20 GeV $<P_{T}<$ 200 GeV is almost 33$\%$ at LHC (Fig.\ref{fig2}).

In conclusion, the photoproductions are very important in the $ep$ deep inelastic collisions at HERA, we extend the photoproduction mechanism to the real photon production in the $pp$ collisionswe at RHIC and LHC energies. Based on the Weizs$\ddot{\mathrm{a}}$cker-Williams approximation,
we derive the high energy photon spectrum from the charged partons, and study the cross section of the photoproduction and resolved photoproduction.
The photon
spectrum from the charged parton depends on the collision energies,
and the photoproductions can produce large $P_{T}$ photons at RHIC and LHC. The numerical results show that the photoproductions
contribute an active modification for the direct photons and fragmentation photons production. The modification of photoproductions is weak in the relatively small $P_{T}$, but becomes important in the large $P_{T}$ region.

This work is supported by the National Natural Science Foundation of
China (10665003 and 11065010).

\begin{thebibliography}{99}




\bibitem{1}
Owens J F 1987 {\it Rev. Mod. Phys.} {\bf 59} 465

\bibitem{12.1}
Peressounko D, the PHENIX Collaboration 2007 {\it Nucl. Phys.} A {\bf 783} 577

\bibitem{19.1}
Fu Y P and Li Y D 2011 {\it Nucl. Phys.} A {\bf 865} 76

\bibitem{19.2}
Fu Y P and Li Y D 2010 {\it Chin. Phys. Lett.} {\bf 27} 101202


\bibitem{19.3}
Fu Y P and Li Y D 2009 {\it Chin. Phys. Lett.} {\bf 26} 111201

\bibitem{19.4}
Fu Y P and Li Y D 2010 {\it Chin. Phys.} C {\bf 34} 186


\bibitem{2}
Field R D 1989 {\it Applications of Perturbative QCD} (New York: Addison-Wesley
Publishing Company) pp 186-195

\bibitem{3}
Kang Z, Oiu J and Vogelsang W 2009 {\it Phys. Rev.} D {\bf 79} 054007
\bibitem{15}
Peitzmann T and Thoma M H 2002 {\it Phys. Rep.} {\bf 364} 175

\bibitem{12}
Turbide S, Gale C, Jeon S and Moore G D 2005 {\it Phys. Rev.} C {\bf 72} 014906

\bibitem{12.01}
Aurenche P, Baier R, Douiri A, Fontannaz M and Schiff D 1987 {\it Nucl. Phys. } B {\bf 286} 553



\bibitem{21}
Nisius R 2000 {\it Phys. Rep.} {\bf 332} 165

\bibitem{22}
Krawczyk M,Zembrzuski A and Staszel M 2001 {\it Phys. Rep.} {\bf 345} 265

\bibitem{23}
Forshaw J R arXiv:hep-ph/9706319



\bibitem{24.1}
Fu Y P and Li Y D 2011 {\it Phys. Rev.} C {\bf 84} 044906


\bibitem{24.2}
Yin R T, Fu Y P and Li Y D 2011 {\it Cent. Eur. J. Phys.} {\bf 9} 1434

\bibitem{24}
Baur G 1998 {\it J. Phys.} G {\bf 24} 1657

\bibitem{25.01}
Eboli O J P, Gonzalez-Garcia M C and Novaes S F 1994 {\it Phys. Rev.} D
{\bf 49} 91

\bibitem{25}
Drees M, Godbole R M, Nowakowski M and Rindani S D 1994 {\it Phys. Rev.} D
{\bf 50} 2335


\bibitem{5}
Gl$\ddot{\mathrm{u}}$ck M, Reya E and Vogt A 1992 {\it Z. Phys.} C {\bf 53} 127



\bibitem{7}
Gl$\ddot{\mathrm{u}}$ck M, Reya E and Vogt A 1992 {\it Phys. Rev.} D {\bf 46} 1973



\bibitem{9.01}
Khachatryan V, the CMS Collaboration 2011 {\it Phys. Rev. Lett.} {\bf 106} 082001


\bibitem{23.01}
Berger E L, Braaten E and Field R D 1984 {\it Nucl. Phys.} B {\bf 239} 52



\end{thebibliography}

\end{document}